\documentclass[pra,twocolumn,amsmath]{revtex4}
\usepackage{graphicx,epsfig}

\begin{document}

\def\ket#1{|#1\rangle}
\def\bra#1{\langle#1|}
\def\av#1{\langle#1\rangle}
\def\myarrow{\mathop{\longrightarrow}}

\title{Effects of random localizing events on matter waves:  formalism and examples}

\author{Julio Gea-Banacloche}
\affiliation{Department of Physics, University of Arkansas, Fayetteville, AR 72701\\
and National Institute of Standards and Technology, Gaithersburg, MD 20899}
\email[]{jgeabana@uark.edu}

\date{\today}

\begin{abstract}
A formalism is introduced to describe a number of physical processes that may break down the coherence of a matter wave over a characteristic length scale $l$.  In a second-quantized description, an appropriate master equation for a set of bosonic ``modes'' (such as atoms in a lattice, in a tight-binding approximation) is derived.  Two kinds of ``localizing processes'' are discussed in some detail and shown to lead to master equations of this general form: spontaneous emission, and modulation by external random potentials.  Some of the dynamical consequences of these processes are considered: in particular, it is shown that they generically lead to a damping of the motion of the matter-wave currents, and may also cause a ``flattening'' of the density distribution of a trapped condensate at rest.    
\end{abstract}

\maketitle

\section{Introduction}

Experiments in which the wavefunction of a material particle (or system of particles) becomes delocalized, in a coherent way, over a relatively large region of space have become increasingly common in recent years, eventually giving rise to the entire subdiscipline of matter-wave physics.  The purpose of this paper is to characterize, and study some of the consequences of a number of physical process that may break this long-range coherence, in particular, for large numbers of particles described in a second-quantized formalism.  These (assumed random) events will be referred to here as ``localizing events'' because, even though they do not necessarily pinpoint the actual position of the particle (as long as they go unobserved), the breakdown of coherence over a scale $l$ results in a density operator that can be consistently interpreted as describing an ensemble of particles localized in regions of space of a size $\sim l$, centered at some unknown point(s) $x_0$, with the $x_0$ ranging over the region occupied by the original coherent wavefunction.  

Traditionally, the study of these types of processes has taken place mostly in the context of research into decoherence and the quantum-to-classical transition (see \cite{joos} for an early review, and \cite{hornberger} for more recent work and further references).  Interferometric experimental observations of decoherence due to particle scattering \cite{hornberger2} and emission of thermal radiation \cite{hackermuller} have recently been reported; these papers also include many references to earlier relevant experiments, such as, e.g., \cite{pfau,chapman,kokorowski}.    

In this paper, a number of physical processes that may result in ``localizing events'' will be described, focusing in particular on systems of cold atoms (including Bose-Einstein condensates) in optical lattices.  A simple master-equation formalism will then be introduced to describe the decoherence induced by these processes in a second-quantized picture, and it will be shown that their effect on the transport properties of the system takes the form of a damping of the matter-wave current.  In one particular case, namely, that of modulation by external random potentials, an expression for the damping is derived that is very similar to one recently proposed \cite{preprint} to explain the damping observed \cite{trey} in certain one-dimensional atomic Bose-Einstein condensates. Finally, the effects of these random localizing events on the density distribution of a trapped condensate at rest are also briefly discussed.

It may perhaps need to be stressed that words such as ``localization'' or ``localizing'' are used in this paper in the sense described above (i.e., breakdown of long-range coherence), and not in the technical sense of, e.g., the Anderson localization transition, or the so-called ``dynamical localization'' phenomenon.  As regards the first of these, the processes considered here are all dynamic in nature, whereas Anderson localization \cite{Anderson}, in its simplest form, concerns the nature of the eigenfunctions of a particle in a disordered but {\em static} potential.  As for dynamical localization, it is a phenomenon that takes place specifically in the quantized versions of classically chaotic systems, which does not really describe most of the systems to be considered here.  That being said, however, it should also be pointed out that some of the systems described in Section V below may, under some circumstances, exhibit classical chaos and/or Anderson localization \cite{Gavish}, so it might be interesting to study the interrelationships between all these phenomena for such systems.  This may be pursued at some later date.

\section{Formalism}
\subsection{Localization by direct wavefunction collapse}
Suppose that we have a single particle in a state described by a wavefunction $\psi(x)$, which we imagine as extending over a relatively large region of space.  A ``localizing event'' is one that would cause the wavefunction to collapse to a region of space of size $l$ (we work in one dimension throughout) around a point $x_0$; the resulting wavepacket may then be described by a wavefunction $f(x-x_0)$, where the function $f(x)$ has width $l$ and is centered at $x=0$.  Clearly, we expect the overall probability of such a collapse to occur to be proportional to
\begin{equation}
p(x_0,f) = \left|\int_{-\infty}^\infty f(x-x_0)^\ast \psi(x) \,dx\right|^2
\label{e1}
\end{equation}
If the location $x_0$ is not known, the final state of the particle will be described by a density operator proportional to
\begin{equation}
\rho(x,x') = \frac{\int_{-\infty}^\infty p(x_0,f)f(x-x_0)^\ast f(x'-x_0)\,dx_0}{\int_{-\infty}^\infty p(x_0,f)\,dx_0}
\label{e2}
\end{equation}
This is properly normalized if $f(x)$ is. We are implicitly assuming that the shape $f$ of the localized wavepacket does not depend on the location $x_0$ of the localizing event, although the formalism could clearly be extended to deal with spatial inhomogeneities.  

At the risk of belaboring the obvious, we emphasize that the state described by (\ref{e2}) may still be ``delocalized'' in the sense that the probability $\rho(x,x)$ to find the particle at point $x$ may still be nonzero over a large region of space (of the order of the original region covered by $\psi(x)$).  What has been mainly lost, in going from $\psi(x)$ to the density operator (\ref{e2}), is the coherence between regions of space separated by a distance greater than $l$.  As these regions can no longer interfere, it is legitimate to interpret (\ref{e2}) as describing a situation in which the \emph{coherent wavefunction} of the particle is \emph{localized} to a much smaller region of space of width $l$, only the precise location of the center of this wavepacket is not known with certainty.

A simple example for which analytical calculations are possible is obtained if $\psi$ and $f$ are both Gaussian wavepackets, $\psi(x) = (2\pi)^{-1/4}\sigma_0^{-1/2}\exp(-x^2/4\sigma_0^2)$ and $f(x) = (2\pi)^{-1/4}l^{-1/2}\exp(-x^2/4l^2)$, where the original $(\Delta x)_0 = \sigma_0$ may be much greater than $l$.  We obtain, from Eq.~(\ref{e1}),
\begin{equation}
p(x_0,f) = \frac{2\sigma_0 l}{\sigma_0^2 + l^2}\,e^{-x_0^2/2(l^2+\sigma_0^2)} \equiv \frac{2\sigma_0 l}{\sigma^2}e^{-x_0^2/2\sigma^2}
\label{e3}
\end{equation}
defining the new $\sigma = (\sigma_0^2+ l^2)^{1/2}$; and, from Eq.~(\ref{e2}),
\begin{align}
\rho(x,x') &= \frac{1}{\sqrt{2\pi}}\,\frac{1}{\sqrt{\sigma^2+l^2}}\,e^{-(x^2+{x'}^2)/4(\sigma^2+l^2)} \notag \\
&\qquad\qquad\times e^{-(x-x')^2\sigma^2/8l^2(\sigma^2+l^2)} \notag \\
&\simeq \frac{1}{\sigma_0\sqrt{2\pi}}\,\,e^{-(x^2+{x'}^2)/4\sigma_0^2} e^{-(x-x')^2 /8l^2} \notag \\
&\simeq \psi^\ast(x) \psi(x') e^{-(x-x')^2 /8l^2}
\label{e4}
\end{align}
where the approximations apply in the case $\sigma_0\gg l$.  Equation (\ref{e4}) shows that, under these circumstances, the region where the particle may be found, as given by $\rho(x,x)$, is substantially the same as before; the difference is in the spatial coherence represented by the off-diagonal terms (those with $x\ne x'$), which is seen to decay with $|x-x'|$ over a region of the order of magnitude of the localizing length $l$.

It is now easy to see that, in the limit in which $l\ll (\Delta x)_0$ (where $(\Delta x)_0$ is the spread of the initial wavefunction) one should generically expect $\rho(x,x')$ to take a form similar to (\ref{e4}).  The reason is that, in Eq.~(\ref{e1}), $f(x-x_0)$ will act as a $\delta$ function, yielding a $p(x_0,f)$ which is proportional to $|\psi(x_0)|^2$ (cf. Eq.~(\ref{e3})); then, in Eq.~(\ref{e2}), $p(x_0,f)$ could be evaluated at $x_0 = x$ (or $x_0=x'$) and factored out of the integral, which then becomes the autocorrelation of $f$, evaluated at $x-x'$.  Thus, we expect, in general,
\begin{align}
\rho(x,x') &\simeq |\psi(x)|^2 g(x-x') \notag \\
&\simeq \psi^\ast(x) \psi(x') g(x-x')
\label{e5}
\end{align}
where $g(x)$ is proportional to the autocorrelation of $f$.  The second approximate equality is because we expect $\psi^\ast(x) \psi(x') \simeq |\psi(x)|^2$ if $|x-x'|\ll (\Delta x)_0$, which will be the case if $|x-x'|$ is restricted to a region of the order of $l$.

\subsection{Localization by momentum kicks}

Consider a particle with wavefunction $\psi(x)$ that receives a momentum ``kick'' of magnitude $\hbar k$ at some point in time.  The new wavefunction will be
\begin{equation}
\psi^\prime_k(x) = e^{i kx}\psi(x)
\label{e6}
\end{equation}
since the operator $e^{ik\hat x}$ is the generator of a momentum translation.  Assume that many different values of $k$ are possible, distributed with a probability $p(k)$. Then the density operator describing the state of the particle will be
\begin{align}
\rho(x,x') &= \int_{-\infty}^\infty p(k) \psi^\prime_k(x)^\ast \psi^\prime_k(x') \,dk \notag\\
&=  \left(\int_{-\infty}^\infty p(k) e^{-i k(x-x')}\,dk \right)\psi(x)^\ast \psi(x')
 \label{e7}
\end{align}
which is clearly of the form (\ref{e5}).  If we search for the $f(x)$ that $p(k)$ is related to, we note that the Fourier transform of a convolution product is the product of the Fourier transforms, so $p(k) \sim |\tilde f(k)|^2$ (where $\tilde f$ is the Fourier transform of $f$). As $p(k)$ is, by construction, positive, its square root is well-defined, and so one can simply take
\begin{equation}
f(x) = \frac{1}{\sqrt{2\pi}} \int_{-\infty}^\infty e^{-ikx} \sqrt{p(k)}\,dk
\label{e8}
\end{equation}
which is already properly normalized. For non-pathological $p(k)$'s, the function $f(x)$ should have a width $l\sim 1/\Delta k$, so we can conclude that random momentum kicks lead to a ``localization'' of the coherence, in the sense of the previous section, within regions whose size $l$ is inversely proportional to $\Delta k$, the range of momenta involved.

\subsection{Representation of the localization process in a second-quantized formalism}

In order to apply these ideas to many-particle systems,  especially bosonic systems, it is convenient to introduce a second-quantized formalism in which the possible states of each individual particle are represented by localized wavepackets, whose width is already of the order of magnitude of the width $l$ associated with the localization processes. This ``Wannier representation'' approach may arise naturally in some important cases; in particular, for cold atoms in optical lattices.  In a single-band description, in which only one type of wavepacket needs to be considered at each site, the general pure state of a single particle can be written as
\begin{equation}
\ket{\psi} = \sum_i C_i \ket{0,\ldots,1,\ldots,0}
\label{e9}
\end{equation}
with amplitudes $C_i$, where the ``$1$'' in the ket is at the $i$-th location and all the other entries are zeros.  In this description, where the ``lattice'' is matched to the localization process, it is easy to see what the action of a localization event is:  a single basis state of the form $\ket{0,\ldots,1,\ldots,0}$ is left unchanged (the particle can only be localized at the $i$-th site in this case, and it is already localized there), but the coherence of a superposition like (\ref{e9}) is totally destroyed, so $\rho$ becomes
\begin{equation}
\rho = \sum_i |C_i|^2 \ket{0,\ldots,1,\ldots,0}\bra{0,\ldots,1,\ldots,0}
\label{e10}
\end{equation}
We can now ask what happens when one has several particles, of the bosonic type (so that more than one of them can be at the same site and with the same wavefunction). We aim at describing the localization process using some appropriate ``jump operator,'' with a view to eventually describing the dynamics of a system exposed to localizing events by a master equation.  The observation made above generalizes to the idea that pure Fock states of the form $\ket{n_1,n_2,\ldots,n_N}$ must be left invariant by a localization event at the $i$-th location:  the event can only happen if $n_i\ne 0$, but other than that the state does not change, since it already involves atoms whose precise locations are known.  This means that the jump operator $L_i$ associated with a localizing event at site $i$ must be a function of $\hat n_i$, the corresponding number occupation operator.  It is also natural to make it \emph{proportional} to $\hat n_i$, since this expresses the fact that a localization event is more likely to occur at a given site, the more particles one actually already has there.  Note that this choice ensures that the single-particle complete loss of coherence, illustrated by the passage from (\ref{e9}) to (\ref{e10}), still takes place: in that case, each jump operator $\hat n_i$ can have only the eigenvalue 0 or 1, so each one of them annihilates all the states except for one, and when the outcomes of all the possible events are added with equal a priori weights one recovers (\ref{e10}).  The novelty now is that a superposition such as, for instance,
\begin{equation}
\ket{\psi} = \frac{1}{\sqrt{2}}(\ket{2,3}+\ket{3,2})
\label{e11}
\end{equation}
is not rendered completely incoherent:  a localizing event at site 1 transforms it into $(2\ket{2,3}+3\ket{3,2})/\sqrt{13}$, and an event at site 2 transforms it into $(3\ket{2,3}+2\ket{3,2})/\sqrt{13}$.  Both events are just as likely, so the final density operator is
\begin{equation}
\rho = \frac{1}{2}\ket{2,3}\bra{2,3} + \frac{1}{2}\ket{3,2}\bra{3,2} +  \frac{6}{13}\ket{2,3}\bra{3,2} + \frac{6}{13}\ket{3,2}\bra{2,3}
\label{e12}
\end{equation}
This is no longer a pure state (since $\rho^2\ne\rho$), so some coherence has been lost, but overall, not much. Intuitively, this is because the localizing processes cannot very well distinguish between the state $\ket{2,3}$ and the state $\ket{3,2}$. On the other hand, the state (\ref{e12}) is not, as will become clear in the following subsection, a terminal state of the evolution of the system:  a sufficiently large number of successive localizing events would eventually destroy completely the coherence of all superpositions such as (\ref{e11}).  Physically, this is consistent with the fact that the coherence in (\ref{e11}) is inseparable from the delocalization of one of the five particles involved:  we can say that, in the state (\ref{e11}), two particles are certainly at site 1, and two others at site 2, but there is a fifth particle that is, in some sense, in both locations at once.  Localizing processes will eventually force it to settle in one site or the other, and at that point the coherence of the superposition will be completely lost.  From this perspective, the reason why it may take many localizing events to achieve this result is that there is a large probability that the particles affected by the individual localizing events should be the ones that are localized already, as opposed to the elusive fifth one.   

\subsection{Master equation treatment}

Having identified the jump operators for the localizing processes in the previous subsection, we can write down a master equation for the density operator of the second-quantized bosonic system on a lattice, in the standard Lindblad form \cite{schenzle},
\begin{align}
\dot\rho &= -r \sum_i\left(L_i^\dagger L_i\rho + \rho L_i^\dagger L_i - 2 L_i\rho L_i^\dagger \right) \notag \\
&= -r \sum_i\left(\hat n_i^2 \rho + \rho \hat n_i^2 - 2 \hat n_i\rho \hat n_i \right)
\label{e13}
\end{align}
where $r$ is some appropriate rate, proportional to the probability per unit time that a localizing event may occur anywhere.  

In subsequent sections (IV and V) it will be shown how equations of the form (\ref{e13}) may, in fact, be derived, more or less from first principles, for some physical situations. For the moment, however, the focus will be on some of the formal consequences of (\ref{e13}).

To begin with, it is obvious that (\ref{e13}) preserves any Fock state, as well as any incoherent superposition of Fock states.  On the other hand, its effects on superpositions are easy to ascertain, because the $\hat n_i$ operators do not mix different Fock states.  Thus, for a general superposition given by a density operator of the form
\begin{equation}
\rho(t) = \sum \rho_{n_1,\ldots,n_N,n'_1,\ldots,n'_N}(t)\ket{n_1,\ldots,n_N}\bra{n'_1,\ldots,n'_N}
\label{e14}
\end{equation}
the equations for the matrix elements $\rho_{n_1,\ldots,n_N,n'_1,\ldots,n'_N}(t)$ all decouple, and we find
\begin{align}
\dot \rho_{n_1,\ldots,n_N,n'_1,\ldots,n'_N}(t) &= -r\left(\sum_i\left(n_i^2-{n_i^\prime}^2 - 2 n_i n_i^\prime\right)\right) \notag\\
&\qquad\times\rho_{n_1,\ldots,n_N,n'_1,\ldots,n'_N}(t)
\label{e15}
\end{align}
Since $(n_i^2-{n_i^\prime}^2 - 2 n_i n_i^\prime)  = (n_i-n_i^\prime)^2 \ge 0$, all the amplitudes $\rho_{n_1,\ldots,n_N,n'_1,\ldots,n'_N}(t)$ eventually decay to zero except those for which, for every $i$, $n_i=n_i^\prime$, that is to say, the diagonal elements.  Hence, as pointed out at the end of the previous subsection, all coherences between different Fock states are eventually destroyed, although some may decay much faster than others, depending on how large the differences $(n_i-n_i^\prime)^2$ are.  

\subsection{Generalizations}

The most important generalization of the formalism is to deal with situations where the localized wavepacket may not be a perfect fit to the ``natural'' lattice for the problem.  In the single-particle case, denoting by $\ket{i}$ the state $\ket{0,\ldots,1,\ldots,0}$ with a 1 in the $i$-th position, it is easy to see that the continuous treatment in Section II.A amounts, in the discrete case, to reducing a state of the form (\ref{e9}) to something of the form
\begin{equation}
\ket{\psi}^\prime_i = \sum_{i'}f(i'-i)\ket{i'}
\label{e16}
\end{equation}
where $f(i)$ is a discrete function of the (negative or positive) integer argument $i$, centered at $i=0$.  The formalism developed in Sections II.C and II.D corresponds to the assumption that the particle's wavepacket collapses to a single lattice site, which is expressed by the choice $f(i)=\delta_{i0}$ ($\delta_{ij}$ is the Kronecker delta symbol).  If the wavepacket spreads over a few lattice sites instead, one can get a result approximately equivalent to (\ref{e16}) by acting on the (single-particle) wavefunction with the operator
\begin{equation}
L_i = \sum_{i'} \hat n_{i'} f(i'-i).
\label{e17}
\end{equation}
This produces $\sum_{i'}f(i'-i)C_{i'}\ket{i'}\simeq C_i \sum_{i'}f(i'-i)\ket{i'}$, provided the original state's coefficients $C_i$ do not change by much over the region where $f(i'-i)$ is nonzero (cf. the remarks at the end of Section II.A and in Section II.B).  The overall factor $C_i$ is later removed by normalization.

It seems, therefore, natural to generalize the concept of ``localization-induced decoherence'' to the many-particle case by adopting (\ref{e17}) for the Lindblad jump operators on the first line of Eq.~(\ref{e13}), and use the resulting master equation to describe the dynamics when the localization processes cover a region somewhat larger than a single lattice site. 

\section{Damping of center-of-mass motion by localizing events.}

Consider a set of $N$ identical particles that may occupy any of a total of $M$ sites in a one-dimensional chain.   The position of the center of mass of this system can be described by the operator
\begin{equation}
\hat x_{CM} = \frac{a}{N}\sum_{j=1}^M j \hat n_j
\label{e18}
\end{equation}
if the lattice spacing is $a$. As is known from standard discretization schemes, such as the Bose-Hubbard model \cite{jaksch,nistrecent}, an appropriate kinetic energy operator for this system is
\begin{equation}
K=- J\sum_{j=1}^M\left(\hat a_j^\dagger\hat a_{j+1} + \hat a_{j+1}^\dagger\hat a_j \right)
\label{e19}
\end{equation}
For convenience we shall assume periodic boundary conditions, so that $M+1 \equiv 1$ (i.e., all the additions of indices will be understood to be performed modulo $M$). The Heisenberg equations of motion, with the Hamiltonian $H=K$, then yield the center of mass velocity
\begin{equation}
\hat v_{CM} = \frac{d\hat x_{CM}}{dt} = \frac{i}{\hbar}[K,\hat x_{CM}] = i\frac{Ja}{ N\hbar} \sum_j\left(a_{j+1}^\dagger a_j- a_j^\dagger a_{j+1}\right)
\label{e20}
\end{equation}
If one applies $[K,\hat v_{CM}]$ again, to calculate the rate of change of the center of mass velocity, one obtains a sum of terms that cancel identically, except for edge effects that are taken care of by the assumed periodic boundary conditions.  This indicates that the Hamiltonian $H=K$ is appropriate for a system of free (and non-interacting) particles.

Imagine now that the system is somehow subjected to ``localizing'' processes, such as those considered in the previous section, that result in a master equation for the many-particle density operator $\rho$ of the form given by (\ref{e13}).  Then we can calculate their effect on the time-derivative of the expectation value of any operator $O$ by the formula $(d/dt)\av{O} = Tr(O\dot\rho)$.  The result for $d\av{\hat x_{CM}}/dt$ is 0, since all the $\hat n_i$ operators commute among themselves.  On the other hand, the result for $d\av{\hat v_{CM}}/dt$ is
\begin{widetext}
\begin{align}
\frac{d}{dt}\av{\hat v_{CM}} &= -r\sum_{i=1}^M\left(Tr(\hat v_{CM}\hat n_i^2 \rho) + Tr(\hat v_{CM}\rho \hat n_i^2) - 2 Tr(\hat v_{CM}\hat n_i\rho \hat n_i) \right) \notag \\
&=-\frac{irJa}{N\hbar}\sum_{i,j}\left\langle (a_{j+1}^\dagger a_j- a_j^\dagger a_{j+1})\hat n_i^2 + \hat n_i^2(a_{j+1}^\dagger a_j- a_j^\dagger a_{j+1})-2 \hat n_i(a_{j+1}^\dagger a_j- a_j^\dagger a_{j+1})\hat n_i \right\rangle \notag \\
&= -\frac{2irJa}{N\hbar}\sum_{j=1}^M \left\langle a_{j+1}^\dagger a_j- a_j^\dagger a_{j+1}\right\rangle \notag \\
&= -2r \av{\hat v_{CM}}
\label{e21}
\end{align}
\end{widetext}
This means that localizing processes of the form considered here result in a damping of the center-of-mass motion of the system, at a rate (i.e., with a damping constant) $2r$.  To some extent, this should not be a surprising result, since we showed in Section II.B that random momentum kicks tend to localize the particle (always in the sense of ``breaking the long-range coherence of the wavefunction'') in a way quite similar to actual random projective position measurements, and it has been known at least since Einstein's days that a particle subjected to random momentum kicks experiences an overall damping of its motion (fluctuation-dissipation theorem).  What we have here is a formal derivation, and unification, of these various results, in a form suitable for use in quantum many-body physics.  A specific example of localization by random momentum kicks is provided in the next Section, by a consideration of atomic spontaneous emission.

Damping of the center of mass motion also results from the more general localization processes discussed in Section II.E, that is, processes that tend to localize the particle over a region larger than one lattice site.  With $L_i$ as given by (\ref{e17}), direct calculation shows that the contribution of the term $L_i^2\rho + \rho L_i^2 -2L_i\rho L_i$ to the time derivative of $\av{a_{j+1}^\dagger a_j}$ is
\begin{align}
&\left\langle a_{j+1}^\dagger a_j L_i^2 + L_i^2 a_{j+1}^\dagger a_j -2 L_i a_{j+1}^\dagger a_jL_i \right\rangle \notag\\
&\qquad= \left(f(j-i)-f(j+1-i)\right)^2 \av{a_{j+1}^\dagger a_j} 
\label{e22}
\end{align}
To calculate the total time derivative, one should sum (\ref{e22}) (and its complex conjugate) over all the values of $i$, which correspond to processes localizing the particles around the respective sites.  When this is done, the prefactor in (\ref{e22}) becomes independent of $j$:
\begin{align}
&\sum_i \left(f(j-i)-f(j+1-i)\right)^2 \notag\\
&\qquad= \sum_i \left(f(-i)-f(1-i)\right)^2 \equiv \bar f^2
\label{e23}
\end{align}
so that now
\begin{equation}
\frac{d}{dt}\av{\hat v_{CM}} = -\bar f^2 r \av{\hat v_{CM}}
\label{e24}
\end{equation}
In general, less localization also means less damping.  If $f(i)=\hat\delta_{i0}$, one has $\bar f^2 = 2$, as in Eq.~(\ref{e21}).  On the other hand, if, for example, one takes the three-point function $f(0) = 1/\sqrt 2$, $f(\pm 1) = 1/2$, Eq.~(\ref{e23}) yields $\bar f^2 = 1/2+2(1/\sqrt  2-1/2)^2 = 0.586$.

\section{Localization through spontaneous emission (light scattering)}
Consider a system of atoms in an optical lattice (as usual, we will consider one dimension only).  If the laser beams used to form the lattice have wavelength $\lambda$, then the lattice spacing is $a=\lambda/2$.  In general, the lasers are far detuned from the atomic transition, but it may still happen from time to time that an atom may absorb a photon from the lattice and reemit it in a random direction.  From the absorption, the atom receives a momentum $\hbar k = \pm 2\pi \hbar/\lambda$, whereas from the spontaneous emission the momentum component along the lattice direction can be anything between $-2\pi \hbar/\lambda$ and $2\pi \hbar/\lambda$.  Overall, therefore, the atom experiences a random momentum kick between  $-4\pi \hbar/\lambda$ and $4\pi \hbar/\lambda$.  Most often, the whole process is referred to as spontaneous emission, but more properly it should be called ``light scattering'' (see \cite{gorlitz} for a discussion of why this distinction may be important; see also \cite{hackermuller,pfau,chapman,kokorowski} for detailed descriptions of the decoherence induced by light scattering by free atoms or molecules).  

In a periodic lattice, the atom's momentum is quantized, and restricted to values $2\pi p\hbar/Ma$, where $M$ is the number of lattice sites and $p$ is an integer between $-M/2$ and $M/2$ (first Brillouin zone).  With $a=\lambda/2$, the range of momenta from $-4\pi \hbar/\lambda$ to $4\pi \hbar/\lambda$ corresponds to $-2\pi \hbar/a$ to $2\pi \hbar/a$, twice the range of the first Brillouin zone.    We shall take momenta that differ by $2\pi \hbar/a$ to be equivalent (i.e., quasimomenta).  For a single particle, we would describe the effect of a random spontaneous emission event on a state of the form (\ref{e9}) as
 \begin{align}
\ket{\psi} &= \sum_jC_j\ket{0,\ldots,1,\ldots,0} \notag\\
& \to \ket{\psi}^\prime_p= \sum_jC_je^{2\pi ip j/M}\ket{0,\ldots,1,\ldots,0}
\label{e25}
\end{align}
where the argument in the exponent is $i(px)/\hbar$, as in Eq.~(\ref{e6}), only $x=ja$ for a particle at site $j$ in the lattice.  

In the many-particle, second-quantized formalism, the natural generalization of (\ref{e25}) is to adopt a ``jump operator'' of the form \cite{javanainen}
\begin{equation}
\tilde L_p = \sum_j e^{2\pi ip j/M} \hat n_j
\label{e26}
\end{equation}
This is most naturally seen by working temporarily in the momentum representation, where the creation and annihilation operators of an atom with momentum $k$ are defined by
\begin{equation}
\hat c_k = \frac{1}{\sqrt M}\sum_{j=1}^M e^{-2\pi i k j/M} \hat a_j 
\label{e27}
\end{equation}
Then if an atom in the momentum state $k$ emits a photon, its momentum changes to $k+p$, and the appropriate operator is simply $c^\dagger_{k+p} c_k$.  The sum of this over all $k$ is
\begin{align}
\sum_{k=-M/2}^{M/2}\hat c^\dagger_{k+p} \hat c_k &= \frac{1}{M}\sum_{k=-M/2}^{M/2}\sum_{j,j'} e^{2\pi i [(k+p) j-kj^\prime]/M}  \hat a^\dagger_j \hat a_{j^\prime} \notag\\
&= \sum_j e^{2\pi ip j/M} \hat n_j \notag\\
&= \tilde L_p 
\label{e28}
\end{align}
We can, therefore, use (\ref{e26}) to write a master equation of the form
\begin{equation}
\dot\rho = -r\frac{1}{M}\sum_p\left(\tilde L_p^\dagger \tilde L_p \rho + \rho \tilde L_p^\dagger \tilde L_p - 2 \tilde L_p \rho \tilde L_p^\dagger \right)
\label{e29}
\end{equation}
where $r$ is the (single-atom) rate of spontaneous emission events, the sum over $p$ is from $-M/2$ to $M/2-1$ (since the quasimomentum $M/2$ is the same as $-M/2$), and the factor $1/M$ gives the probability that, in each event, the momentum kick received by the atom may be precisely $p$.  This uniform probability distribution for $p$ has been adopted for simplicity, as it immediately leads, via carrying out the sum over $p$, to 
\begin{equation}
\dot\rho = -r \sum_i\left(\hat n_i^2 \rho + \rho \hat n_i^2 - 2 \hat n_i\rho \hat n_i \right)
\label{e30}
\end{equation}
which is to say, the same as the simple master equation (\ref{e13}) derived in Section II.D.  It follows that, under these conditions, spontaneous emission would cause damping of the center of mass motion with damping constant $2r$, as in Section III.

If the probability distribution $g(p)$ of the momentum kicks is not uniform, one would obtain a somewhat different master equation:
\begin{equation}
\dot\rho = -r \sum_{i,j}\tilde g(i-j)\left(\hat n_i \hat n_j \rho + \rho \hat n_i \hat n_j - 2 \hat n_i\rho \hat n_j \right)
\label{e31}
\end{equation}
where $\tilde g(i-j) = \sum_p g(p) e^{2\pi i p(i-j)/M}$ is a discrete Fourier transform of $g(p)$.  Note that this will be of the same form as the generalization discussed at the end of Section III, and based on the jump operator (\ref{e17}), provided one has a function $f$ such that
\begin{equation}
\tilde g(j-j') = \sum_{i} f(j-i)f(j'-i)
\label{e32}
\end{equation}
Ignoring possible complications arising from the finite size of the lattice, we see that Eq.~(\ref{e32}) may be satisfied by taking $f(j)$ to equal the Fourier transform of $\sqrt{g(p)}$ (cf. Section II.D and Eq.~(\ref{e8})).  Thus, in this more general case also, and provided that the range of probable momentum kicks is sufficiently large, one obtains the effective localization of the wavefunction and the attendant damping of the center of mass motion, as discussed in Section III.

The results in this Section are consistent with the intuitive notion that a spontaneously emitted photon carries information on the position of the atom, and therefore tends to partially collapse its wavefunction.  On the other hand, a simple picture of the atoms in a 1-D Bose-Einstein condensate (such as the ``tubes'' to be discussed in the next section) as ``billiard balls'' might suggest that a spontaneous emission event simply gives a momentum kick that is transmitted down the chain of identical atoms and eventually leads to no other consequence that the ejection (from the condensate, if not from the actual physical lattice) of the last atom in the chain.  Here we see that, in fact, the partial localization, and attendant decoherence of the many-particle wavefunction, caused by the emission event also results in an overall damping of the center of mass velocity for a condensate in motion.  

In a far-detuned optical lattice the ``potential depth'' $V$ is proportional to $\Omega^2/\Delta$, where $\Delta$ is the detuning and $\Omega$ the Rabi frequency, whereas the light scattering rate $r$ is proportional to $(\Omega/\Delta)^2\Gamma \sim V\Gamma/\Delta$, where $1/\Gamma$ is the lifetime of the excited state \cite{jessen}.  Since $r$ is typically very small in current experiments, the motional damping predicted here may not have been observed yet, but it should be a relatively easy matter to verify it experimentally, in a low-damping, pseudo 1-D geometry (such as the so-called ``pancakes'') by reducing the detuning $\Delta$.  If this is done while increasing $\Omega$, so as to keep the lattice depth $V$ constant, one should observe a motional damping $\gamma \sim 1/\Delta$, regardless of other factors (such as the geometric considerations that might determine $g(p)$ above). Note that this particular damping mechanism is, in fact, independent of the lattice depth (at constant scattering rate).

\section{Localization caused by randomly-fluctuating potentials}

\subsection{Formalism}

At least under some circumstances, random, fluctuating potentials acting on a system of bosons may lead to master equations of the forms discussed in previous sections.  Consider a Hamiltonian with a term
\begin{equation}
H=\sum_i \hat n_i V_i(t) +\ldots
\label{5.1}
\end{equation}
The functions $V_i$ represent varying, random potentials acting at different sites on the lattice. They will be taken to have zero average and be characterized by spatial and temporal correlations that will be more precisely considered in what follows.

The equation of motion for the density operator,
 \begin{equation}
\dot\rho =-\frac{i}{\hbar}[H,\rho]
\label{5.2}
\end{equation}
can be formally integrated once, over a short time interval $[t-\Delta t, t]$, and iterated, to yield
 \begin{equation}
\dot\rho =-\frac{i}{\hbar}[H(t),\rho(t-\Delta t)]-\frac{1}{\hbar^2}\int_{t-\Delta t}^t[H(t),[H(t'),\rho(t')]]\,dt'
\label{5.3}
\end{equation}
We are interested in the particular contributions to this equation arising from the term (\ref{5.1}).  To this end we introduce a modified density operator which involves an average over different realizations of the stochastic process $V_i$.  The first term in (\ref{5.3}) then vanishes, because we do not expect the value of the external potential at $t$ to be correlated in any particular way to the value of $\rho$ at any time $t'\le t$.  This leaves
 \begin{align}
\dot\rho \simeq -\frac{1}{\hbar^2}\sum_{i,j}\int_{t-\Delta t}^t &\av{V_i(t)V_j(t')}\bigl(\hat n_i \hat n_j\rho(t') + \rho(t')\hat n_i \hat n_j \notag\\
&- \hat n_i \rho(t')\hat n_j - \hat n_j \rho(t')\hat n_i  \bigr)\,dt'
\label{5.4}
\end{align}
Note that the above does involve a sort of Markov assumption, since $\rho(t')$ and $V_i(t)$ could certainly become correlated in general: for instance, $\rho(t')$ could depend on $V_i(t^{\prime\prime})$ at an earlier time $t^{\prime\prime}$, and $V_i(t)$ and $V_i(t^{\prime\prime})$ could be correlated.  Perhaps a better way to justify the approximation (\ref{5.4}) is to simply assume that $\rho(t)$ does not change very much over the time scale over which $V_i(t)$ becomes uncorrelated.  Under those conditions, we get the approximate equation of motion
 \begin{equation}
\dot\rho \simeq -\frac{\tau_c}{\hbar^2}\sum_{i,j}\av{V_i V_j}\left(\hat n_i \hat n_j\rho + \rho\hat n_i \hat n_j - 2\hat n_i \rho\hat n_j \right)
\label{5.5}
\end{equation}
where now everything is evaluated at time $t$, $\tau_c$ is a characteristic correlation time for the fluctuating potential, and $\av{V_i V_j}$ is an equal-time average, i.e., we assume the space-time correlation function of the fluctuations factorizes into a space part and a time part, as in, for instance, $\av{V_i(t)V_j(t')} = \av{V_i V_j} e^{-|t-t'|/\tau_c}$.  

Equation (\ref{5.5}) is already formally reducible to cases considered earlier, at least for a homogeneous system (see Eq.~(\ref{e31}), in the previous section, and the discussion following it), but for later purposes it is useful to consider explicitly what happens if we introduce at this point the momentum components of the fluctuating potentials $V_i$: let
\begin{equation}
V_j = \frac{1}{\sqrt M}\sum_{k=0}^{M-1} \tilde V_k e^{2\pi i j k/M} 
\label{5.6}
\end{equation}
For a homogeneous system, we want $\av{V_j^2}$ to be independent of $j$, and $\av{V_jV_{j'}}$ to depend only on $j-j'$.  This means we must assume $\av{ \tilde V_k \tilde V_{k'}} =\av{| \tilde V_k|^2} \delta_{k,-k'}$ (it is understood that, because of the lattice periodicity, $\tilde V_{-k} = \tilde V_{M-k}$).  Then we have
\begin{equation}
\av{V_j V_{j'}} =   \frac{1}{M}\sum_k \av{ |\tilde V_k|^2} e^{2\pi i( j -j')k/M} 
\label{5.7}
\end{equation}
Introducing
\begin{equation}
(\av{|\tilde V_k|^2})^{1/2} = \frac{1}{\sqrt M}\sum g_l e^{-2\pi i k l/M} = \frac{1}{\sqrt M}\sum g_l^\ast e^{2\pi i k l/M}
\label{5.8}
\end{equation}
we find
\begin{align}
\sum_k \av{|\tilde V_k|^2}e^{2\pi i( j -j')k/M} &= \frac{1}{M}\sum_{l, l'} g_l g_{l'}^\ast \sum_k e^{2\pi ik(j-j'-l+l')/M} \notag \\
&=\sum_{l,l'}g_l g_{l'}^\ast \delta_{l',l-j+j'} \notag \\
&=\sum_l g_{l-j'} g_{l-j}^\ast \notag \\
&=\sum_l g_{j'-l}^\ast g_{j-l} 
\label{5.8a}
\end{align}
and (\ref{5.5}) becomes an equation of the form
\begin{equation}
\dot\rho = -r\sum_l\left(L_l^\dagger L_l\rho + \rho L_l^\dagger L_l - 2 L_l \rho L_l^\dagger \right)
\label{5.9}
\end{equation}
with $r=-\tau_c/\hbar^2 M$, and the Lindblad operators 
\begin{equation}
L_l = \sum_jg_{j-l}\, \hat n_j
\label{5.10}
\end{equation}
just like the ones discussed in Section II.E (compare Eq.~(\ref{e17})), only the discrete function $f(i)$ is now complex.  The damping analysis in Eqs.~(\ref{e22})--(\ref{e23}) carries through, with the result
\begin{align}
\gamma &= r\sum_l\left|g_{-l} - g_{1-l}\right|^2 \notag \\
&= \frac{\tau_c}{\hbar^2 M} \sum_l \left|\frac{1}{\sqrt M}\sum_k (\av{|\tilde V_k|^2})^{1/2} e^{-2\pi i k l/M}\left(1-e^{2\pi i k/M}\right) \right|^2 \notag \\
&= \frac{4\tau_c}{\hbar^2 M} \sum_k \av{|\tilde V_k|^2} \sin^2\left(\frac{\pi k}{M}\right) 
\label{5.11}
\end{align}
in terms of the momentum components, $\tilde V_k$, of the fluctuating potential $V_i$.  By undoing the Fourier transformation (\ref{5.6}), this can also be rewritten in terms of the original $V_i$:
\begin{equation}
\gamma = \frac{\tau_c}{\hbar^2 M}\sum_i \left\langle(V_i - V_{i+1})^2\right\rangle
\label{5.11b}
\end{equation}  

\subsection{Application: damping of oscillations in 1-D Bose-Einstein condensates}

The results in the previous subsection may be used to provide a heuristic explanation of the damping observed in some recent experiments \cite{trey} on atomic Bose-Einstein condensates confined to an essentially one-dimensional geometry (a narrow ``tube''), with a relatively weak periodic lattice in the direction of the motion.  It is known that the mean-field (Gross-Pitaevsky) description of a  BEC in a 1-D lattice exhibits a dynamical instability when the wavenumber associated with the supercurrent flow becomes larger than $\pi/2a$ (where $a$ is the lattice spacing) \cite{wu}.  Although in the experiments of \cite{trey} the average velocity of the condensate was kept well below the instability point, the large noncondensate fraction in these tightly-confined systems includes a non-negligible fraction of atoms with momenta larger than $\hbar\pi/2a$.  It has been argued by several groups \cite{polkovnikov,ruostekoski} that this fraction of atoms is ultimately responsible for the damping observed in the experiments \cite{note}.

A heuristic explanation might proceed as follows.  The displacement of the lattice in the experiments turns the ground-state ``vacuum'' fluctuations into ``real'' excitations, without, however, substantially changing the fraction of atoms with momenta above the critical value (since the displacement is quite small).  The chaotic motion of this small fraction of atoms is seen by the rest as producing a random ``external'' potential, through the atom-atom interaction term (the term $(U/2)\sum_i \hat n_i^2$ in the Bose-Hubbard Hamiltonian).  More specifically, a density fluctuation $\delta n_i$ at position $i$ is seen as a fluctuating potential $V_i \equiv U\delta n_i$ by the bulk of the atoms at that site.  We can, therefore, immediately apply the results of the previous subsection to conclude that the center-of-mass motion of the condensate will be damped as $d\av{\hat v_{CM}}/dt = -2\gamma'\av{\hat v_{CM}}$, with 
\begin{equation}
\gamma' = \frac{2\tau_c U^2}{\hbar^2 M} \sum_k \av{|\tilde{\delta n}_k|^2} \sin^2\left(\frac{\pi k}{M}\right)
\label{5.12}
\end{equation}
where $\tau_c$ is a characteristic correlation time for the fluctuations, and $\tilde{\delta n}_k$ is given by the spatial Fourier transform of the fluctuations.  (The factor of 2 in the damping $\gamma' = \gamma/2$ has been introduced to facilitate comparison with the experimental results as well as with previous theoretical estimates.) As indicated at the end of the previous subsection, this can also be rewritten as
\begin{equation}
\gamma' = \frac{\tau_c U^2}{2\hbar^2 M} \sum_i \left\langle (\delta n_i - \delta n_{i+1} )^2\right\rangle
\label{5.13}
\end{equation}

The formula (\ref{5.13}) is very similar to one introduced in \cite{preprint}, but appears to differ from it by a factor of 4, if one should simply identify the $\delta n_i$ above with the noncondensate site densities $\tilde n_i$ considered in \cite{preprint}.  This is mostly because in \cite{preprint} the operator nature of the fluctuation field was preserved for more steps in the calculation, until at the end certain products of bosonic operators were factorized using a standard ansatz that naturally brings about a factor of 2 (e.g., $\av{a^\dagger_j a^\dagger _j a_j a_{j+1}} \simeq 2 \av{a^\dagger_j a_j}\av{a^\dagger _j a_{j+1}}$).  This suggests that in trying to use (\ref{5.12}) or (\ref{5.13}) for quantitative predictions, one may have to define the magnitude of the ``external field'' $\delta n_i$ in a somewhat ad hoc way.

The range of momenta, $\hbar\pi/2a \le |p| \le \hbar\pi/a$, in the ``mean-field unstable'' region corresponds to $M/4\le k\le 3 M/4$ in (\ref{5.12}), and this is centered on the maximum of the trigonometric function.  This suggests that we could obtain an order-of-magnitude estimate of the expected damping by making the replacement    
\begin{equation}
\frac{1}{M} \sum_k \av{|\tilde{\delta n}_k|^2} \sin^2\left(\frac{\pi k}{M}\right) \sim \left(\frac{N n_\text{high}}{M}\right)^2
\label{5.14}
\end{equation}
where $n_\text{high}$ is the fraction of the $N$ atoms in the tube with momenta higher than $\hbar\pi/2a$.  This is a quantity that was calculated, for the experimental parameters, in \cite{preprint}.   We may further assume that $\tau_c \sim \hbar/J$, since $J/\hbar$ is the ``hopping rate'' between neighboring sites, and one may expect this to be the typical rate at which local density fluctuations would decay.  This estimate,
\begin{equation}
\gamma' \sim \frac{2U^2}{\hbar J} \left(\frac{N n_\text{high}}{M}\right)^2
\label{e45}
\end{equation}
is compared in Figure 1 to the experimental data, with remarkably good agreement over a range of values of the lattice depth $V$ where $\gamma'$ changes by more than two orders of magnitude.  (The formula used to fit the high-momentum data in Fig.~1 of \cite{preprint} is $n_\text{high} = 0.01 + 0.018 V + 0.0019 V^2$; see \cite{preprint,nistrecent} for details on how to relate $U$ and $J$ to the lattice depth $V$. The total number of atoms $N=80$ and the number of lattice sites $M = 60$.) 

\begin{figure}
\includegraphics{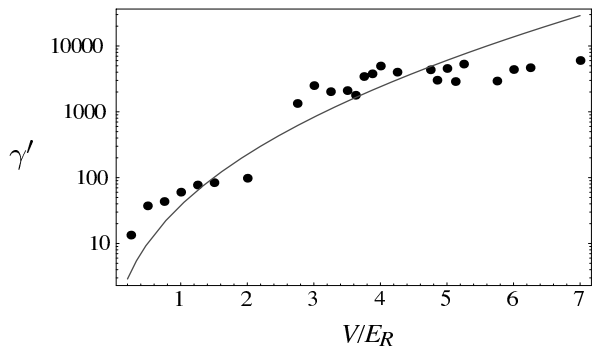}
\caption{\label{fig2} Comparison of the experimentally observed damping (dots) and the estimate (\ref{e45}) (line), as a function of lattice depth $V$ in units of the recoil energy $E_R$. The vertical axis is in units of rad/s.}
\end{figure}

The agreement shown in Fig.~1 suggests that this heuristic approach does capture the basic physics, at least in a qualitative way, for lattices that are not too deep. (For the deeper lattices, where the damping seen in Fig.~1 appears to saturate, a very different pseudo-fermionization model developed in \cite{guido} is found to yield excellent quantitative agreement with the experiments.)  A much more careful treatment of the present model, with a better factorization ansatz, and $\tau_c$ and $\left\langle (\delta n_i - \delta n_{i+1} )^2\right\rangle$ calculated numerically from mean-field theory, can be found in \cite{preprint}.  

It should be noted, however, that there are obvious limitations to the idea of treating the fluctuating field due to the atoms themselves as an equivalent ``external field.'' As will be shown in the next section, the localization process described by a master equation such as (\ref{e13}) is equivalent to a sort of heating, that would eventually destroy completely the coherence of the wavefunction on all length scales; in BEC terms, it would lead to total depletion of the condensate. In reality, however, the system of atoms in the experiment \cite{trey} is, to a good approximation, closed and Hamiltonian, and the only external energy input it receives comes from the initial displacement of the trap.  One would expect, therefore, that the modeling by an external fluctuating field would become inadequate once that small initial energy became ``thermalized'' (if not earlier).

\section{Effects on the static density distribution}

Although most of this paper has been concerned with the effects of a master equation of the form (\ref{e13}) on the system's dynamics, the processes leading to (\ref{e13}) could also have observable effects on the atomic density distribution for a system at rest in a trap (inhomogeneous potential).  This brief section takes a look at these effects.

The Hamiltonian for a system in a trap potential given by the site function $V(i)$ could be written in the form
\begin{equation}
H=\sum_{j}V(j)\hat n_j -J\sum_j \left(\hat a_j^\dagger \hat a_{j+1} + \hat a_{j+1}^\dagger \hat a_j \right)
\label{e50}
\end{equation}
Now we are not assuming a finite number of sites nor periodic boundary conditions: the potential $V(i)$ will determine which sites are, in fact, occupied.  Direct interactions between the particles have been neglected for simplicity; a full treatment including interactions would be quite complicated and beyond the scope of the present paper.

The full equation of motion for the density operator $\rho$ will be 
\begin{equation}
\frac{d\rho}{dt} = -\frac{i}{\hbar}[H,\rho] -r \sum_i\left(\hat n_i^2 \rho + \rho \hat n_i^2 - 2 \hat n_i\rho \hat n_i \right)
\label{e51}
\end{equation}
This leads to a linear system of equations for the expectation values $\av{\hat a^\dagger_i \hat a_j}$ that make up the so-called ``single-particle density matrix'':
\begin{align}
\frac{d}{dt}\av{\hat a^\dagger_i \hat a_j} &= -2r(1-\delta_{ij})\av{\hat a^\dagger_i \hat a_j} +i(V(i)-V(j))\av{\hat a^\dagger_i \hat a_j} \notag \\
&+ iJ\left(\av{\hat a^\dagger_{i+1} \hat a_j}+\av{\hat a^\dagger_{i-1} \hat a_j} - \av{\hat a^\dagger_i \hat a_{j+1}}-\av{\hat a^\dagger_i \hat a_{j-1}}\right)
\label{e52}
\end{align}
In the absence of the damping term $r$, the system would most likely be found in the ground state of the Hamiltonian $H$, where all the expectation values $\av{\hat a^\dagger_i \hat a_j}$ are proportional to the products $C_i^\ast C_j$ of a single-particle wavefunction of the form (\ref{e9}).  With $r\ne0$, however, the system (\ref{e52}) becomes, formally, dissipative, and must evolve towards a final steady state in which all the $\av{\hat a^\dagger_i \hat a_j}$ with $i\ne j$ are zero, and all the $\av{\hat n_j}$ are equal (since the right-hand side of Eq.~(\ref{e52}) involves the differences $\av{\hat n_{j\pm1}}-\av{\hat n_j}$ when $i=j\pm1$).  That is to say, the ultimate steady state of (\ref{e52}) for nonzero $r$ is a ``fully-depleted condensate,'' spatially completely homogeneous, even in the presence of a confining, inhomogeneous potential  

The explanation for this last result must be sought in the fact that the processes responsible for (\ref{e13}) represent a sort of heating, and if they are allowed to continue indefinitely they will eventually provide the system with enough energy to render the confinement potential $V$ irrelevant.  In practice, one expects that the description (\ref{e13}) will cease being even approximately valid long before a totally homogeneous state is reached; but for short times (assuming that one could ``turn on'' the $r$ term in Eqs.~(\ref{e52}) at will, by, for instance, reducing the detuning and bringing up the spontaneous emission rate, as discussed in Section IV), the distinctive effect of the processes (\ref{e13}) on an inhomogeneous system at rest will be a {\it flattening} of the spatial distribution.  

\begin{figure}
\includegraphics[scale=0.65]{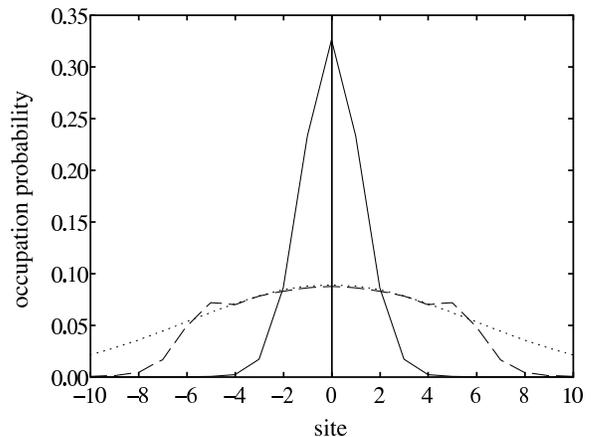}
\caption{\label{newfig2} Evolution of the density distribution for a particle in a parabolic potential $V(j)=\Omega j^2$ with $\Omega/J = 0.1$.  Solid line: ground state density distribution, $r=0$.  Dashed line: density distribution after $t=4\hbar/J$, assuming $r=0.5 J/\hbar$.  Dotted line: a Gaussian fit to the central portion of the dashed line.}
\end{figure}

This effect is illustrated in Fig.~2 for a system of atoms in a parabolic trap.  The potential is chosen so that the initial ground state (with $r=0$) is given by the solid line, which is approximately Gaussian (as one would expect for the ground state of a harmonic oscillator).  Then the $r$ term is turned on, and after a time equal to $8/r$ the new density distribution is given by the dashed line.  To see that the effect is truly a flattening, and not just a broadening, a Gaussian with the same curvature at the center has also been plotted, for reference (dotted curve): it clearly is much broader than the actual solution.

There are a number of ways in which this effect might be useful.  For instance, it might be possible, by watching the shape of the distribution change, to estimate the rate at which ``localizing processes'' of the type considered here are taking place in a given system.  Another intriguing possibility might be to deliberately, and temporarily, ``turn up'' the value of $r$ before driving a trapped BEC through the Mott insulator transition, and see if this results (necessarily after further cooling) in a more uniform filling of the lattice in the insulator phase---a sort of annealing approach.  More detailed calculations, including interaction terms, would be necessary to really assess this possibility.

\section{Conclusions}

In this paper, a formalism has been developed to describe the loss of long-range coherence of a many-particle wavefunction due to random ``localizing'' events, including the physically relevant case of random momentum kicks.  The resulting master equation formalism is especially appropriate to the study of cold atoms in optical lattices.  It has been shown that these random localizing events lead to a damping of the center of mass motion of the system, and several types of processes have been identified as producing these effects:  in particular, spontaneous emission (or, more precisely, light scattering), and random external potentials with the appropriate correlations.  The latter have been used to provide a heuristic model of the recently-observed strong damping of the motion of one-dimensional BECs in a relatively weak lattice.  Finally, the signature that these processes might have on the density distribution of a condensate at rest has also been considered.  

The relatively simple calculations presented here could clearly be extended in several ways, in order to deal with more realistic systems: for instance, using the detailed momentum distribution of scattered photons in spontaneous emission, specific examples of random external potentials, the effects of interactions$\ldots$  If the formalism introduced here is found to be sufficiently promising for dealing with real-world problems, these issues will almost certainly be explored in further work.

\begin{acknowledgments}
I am grateful to NIST for hospitality, and to Ana Mar\'\i a Rey, Carl Williams, Guido Pupillo, and the members of the experimental team of Bill Phillips and Trey Porto for many useful discussions over the past year.  This research has been partly supported by the Army Research Office.  
\end{acknowledgments}


\begin{thebibliography}
\bibitem{note} Note, however, that the application of the dynamical instability concept, originally derived for the Gross-Pitaevski equation in a homogeneous lattice, to the noncondensate atoms in a combination of a periodic lattice and parabolic trapping potential, is rather nontrivial, and, in particular, one cannot assume a sharp transition to unstable motion at the quasimomentum $\hbar\pi/2a$.  Nonetheless, this will be retained as a convenient order of magnitude for the estimates that follow.
\end{thebibliography}
\end{document}